\documentclass[preprint,preprintnumbers,amsmath,amssymb]{revtex4}
\usepackage{amsmath,amssymb}
\usepackage{graphicx, subfigure} \usepackage{dcolumn} \usepackage{bm}
\usepackage{multirow}
\usepackage{color}

\def\vector#1{{\bf #1}}
\def\rcrystal{{\left( \vector{a}^{-1} \vector{r} \right)}}
\def\ngridsubi{{N_{{\rm grid},i}}} \def\beq{\begin{eqnarray}}
	\def\eeq{\end{eqnarray}}  \def\Rvec{{\bf R}}
	\def\rvec{{\bf r}} \def\kvec{{\bf k}} \def\xt{\tilde{x}}
	\def\yt{\tilde{y}} \def\zt{\tilde{z}} \newcommand{\rightbrace}[2]
	{\multirow{#1}{#2}{%
	$\left.{\vcenter{\hsize=0pt\vrule height #1\baselineskip width 0pt}}%
	\right\}$}} \newcommand{\leftbrace}[2]
	{\multirow{#1}{#2}{
	#1\baselineskip width 0pt}}

\begin{document}

\title{Comparison of polynomial approximations to speed up planewave-based
quantum Monte Carlo calculations}

\author{William D. Parker$^1$} \author{C. J. Umrigar$^2$} \author{Dario Alf\`e$^3$} \author{Richard G. Hennig$^4$} \author{John W. Wilkins$^1$} \affiliation{ $^1$Department of Physics, Ohio State University, Columbus, OH 43210, USA\\ $^2$Laboratory of Atomic and Solid State Physics, Cornell University, Ithaca, NY 14853, USA\\ $^3$Department of Earth Sciences and Department of Physics and Astronomy, University College London, London, WC1E 6BT, UK\\ $^4$Department of Material Science and Engineering, University of Florida, Gainesville, FL 32611, USA }

\date{\today}

\begin{abstract}
  The computational cost of quantum Monte Carlo (QMC) calculations of realistic periodic systems depends strongly on the method of storing and evaluating the many-particle wave function.  Previous work [A.~J.~Williamson {\it et al.}, Phys. Rev. Lett. {\bf 87}, 246406 (2001); D.~Alf\`e and M.~J.~Gillan, Phys. Rev. B, {\bf 70}, 161101 (2004)] has demonstrated the reduction of the $O(N^3)$ cost of evaluating the Slater determinant with planewaves to $O(N^2)$ using localized basis functions.  We compare four polynomial approximations as basis functions -- interpolating Lagrange polynomials, interpolating piecewise-polynomial-form (pp-) splines, and basis-form (B-) splines (interpolating and smoothing).  All these basis functions provide a similar speedup relative to the planewave basis.  The pp-splines have eight times the memory requirement of the other methods. To test the accuracy of the basis functions, we apply them to the ground state structures of Si, Al, and MgO. The polynomial approximations differ in accuracy most strongly for MgO and smoothing B-splines most closely reproduce the planewave value for of the variational Monte Carlo energy.  Using separate approximations for the Laplacian of the orbitals increases the accuracy sufficiently to justify the increased memory requirement, making smoothing B-splines, with separate approximation for the Laplacian, the preferred choice for approximating planewave-represented orbitals in QMC calculations.
\end{abstract}

\maketitle

\section{Introduction} \label{sec:intro}

Quantum Monte Carlo (QMC) methods can accurately calculate the electronic structure of real materials\cite{QMCreview,NigUmr-BOOK-99,Kolorenc11}. The two most commonly used QMC methods for zero temperature calculations are variational Monte Carlo (VMC), which can compute expectation values of operators for optimized trial wave functions, and fixed-node diffusion Monte Carlo (DMC), which improves upon VMC results by using the imaginary-time evolution operator to project the trial wave function onto the ground state subject to the fixed-node boundary condition\cite{Anderson77}. QMC has been used to calculate a variety of properties such as cohesive energies, defect formation energies, and phase transition pressures\cite{Yao96, Gaudoin02, Hood03, Maezono03, Needs03, Alfe04b, Alfe05a, Alfe05b, Drummond06, Batista06, Maezono07, Pozzo08, Kolorenc08, Sola09, Hennig10, Driver10, Maezono10, Parker11, Abbasnejad12, Schwarz12, Hood12, Azadi13, Ertekin13, Shulenburger13, Chen14, Benali14, Azadi14, Foyevtsova14}. The accuracy is limited mostly by the fixed-node approximation\cite{Anderson77, Parker11} and the computational power required to reduce statistical uncertainty (the subject of this paper).

Minimizing the time for a QMC calculation of a property (e.g., energy) to a given statistical accuracy requires minimizing the evaluation cost of the orbitals -- used in the trial wave function $\Psi({\bf R})$ -- at each sampling point ${\bf R}$ of the electron coordinates.  The QMC energy, $E_{\rm QMC}$, is a weighted average of the {\it local energy},
\begin{equation}
  E_L(\Rvec) = \frac{H \Psi(\Rvec)}{\Psi(\Rvec)},
\end{equation}
at $N_{\rm MC}$ stochastically-chosen configurations:
\begin{equation} E_{\rm QMC} = \frac{1}{N_{\rm MC}}
	\sum_{i=1}^{N_{\rm MC}} w_i E_L(\Rvec_i).  \label{qmc_energy}
\end{equation}

The statistical uncertainty in $E_{\rm QMC}$ is proportional to $1/\sqrt{N_{\rm MC}}$.  Thus, repeated evaluation of the wave function $\Psi(\Rvec)$ and the Hamiltonian $H$ acting on the wave function, which requires both the wave function and its first and second derivatives, reduces the statistical uncertainty in the calculated property.  The root-mean-square fluctuation of the local energy in VMC
\begin{equation}
  \sigma_{\rm VMC} =
  \sqrt{\frac{1}{N_{\rm MC}} \sum_{i=1}^{N_{\rm MC}} (E_L(\Rvec_i)-E_{\rm
      VMC})^2} \label{qmc_sigma}
\end{equation}
indicates the quality of $\Psi(\Rvec)$ because the individual local energies equal the average when $\Psi(\Rvec)$ is an exact eigenfunction of $H$.  QMC simulations frequently use the Slater-Jastrow form of the wave function\cite{QMCreview}, $\Psi(\Rvec) = J(\Rvec) D(\Rvec)$, where $J(\Rvec)$ is a Jastrow factor\cite{JastrowFactor} (in this work, a simple electron-electron Jastrow with no free parameters is used to impose the electron-electron cusp condition) and $D(\Rvec)$ is a Slater determinant\cite{SlaterDeterminant} of single-particle orbitals.

The orbitals used in QMC wave functions typically come from density-functional or Hartree-Fock calculations and, in periodic systems, are Bloch functions of the form
\begin{equation}
  \phi_{n \kvec} (\rvec) = u_{n \kvec}(\rvec) e^{i\kvec\cdot\rvec},
\end{equation}
where $u_{n \kvec}(\rvec)$ has the periodicity of the crystal lattice, $n$ is the band index, and $\kvec$ the crystal momentum.  The periodic function, $u_{n \kvec}(\rvec)$, is represented by a linear combination of basis functions.  Frequently QMC calculations are performed using simulation cells larger than the primitive cell to reduce Coulomb finite-size errors.  However, since $u_{n \kvec}(\rvec)$ is periodic in the primitive cell, representing it by basis-function expansions in just the primitive cell is sufficient to simulate larger cells.

The computational cost per $N$-electron Monte Carlo move of evaluating the Slater determinant is ${\it O}(N^3)$, when spatially-extended basis functions are used to represent the orbitals, since $N$ orbitals are evaluated for each of the $N$ electrons, and each orbital is a sum over $O(N)$ basis functions.  Spatially-localized basis functions avoid the linear scaling of the number of basis functions with system size since only those basis functions that are non-zero at a given point contribute to the wave function value at that point, resulting in ${\it O}(N^2)$ scaling.

Planewaves, despite their undesirable scaling, are a popular choice for basis functions for the density-functional and Hartree-Fock methods because of their desirable analytic properties.  The advantage of a planewave representation is that planewaves form an orthogonal basis, and, in the infinite sum, a complete single-particle basis.  Thus, adding more planewaves to a truncated basis (as is always used in practice) systematically improves the wave function representation towards the infinite single-particle basis limit.  The energy of the highest frequency planewave included in the sum, the cutoff energy $E_{\rm cut} = \hbar^2 {G_{\rm cut}}^2 / 2\,m_{\rm e}$, 
characterizes a given truncated planewave basis by setting the smallest length scale about which the wave
function has information. Thus, a planewave-based orbital $\phi_{\rm PW}$ is a sum over each planewave ${\bf G}$ below the cutoff multiplying a real- or complex-valued coefficient $c_{{\bf G} n {\bf k}}$ unique to that planewave, the band index $n$, and the crystal momentum ${\bf k}$:

\begin{equation}
  \label{eq:pw_orbital}
  u_{n {\bf k}}({\bf r}) = \phi_{\rm PW}(\bf{r}) = \sum_{\bf G} c_{{\bf G} n {\bf k}} \exp(\imath {\bf G}\cdot {\bf r}).
\end{equation}

Williamson {\it et al.}\cite{Williamson01} first applied the pp-form spline interpolation method to approximate planewave-based orbitals by localized basis functions in QMC calculations.  They report an $O(N)$ reduction in the time scaling.  Alf\`e and Gillan\cite{Alfe04} introducing the related method of B-spline approximation in QMC, report significant reduction in the calculation time while maintaining planewave-level accuracy.

This work compares the three methods previously applied to QMC (pp-splines\cite{Williamson01}, interpolating B-splines\cite{Esler_unpub}, and smoothing B-splines\cite{Hernandez97, Alfe04}) with a fourth method (Lagrange polynomials) originally implemented by one of us in QMC but heretofore unpublished.  Section~\ref{sec:methods} introduces, compares and contrasts the four methods.  Section~\ref{sec:accuracy} compares the accuracy of the polynomial methods in reproducing the QMC energies and fluctuations in the local energy relative to the corresponding values from the planewave expansion.  Section~\ref{sec:accuracy} also studies whether it is advantageous to construct separate approximations for the gradient and the Laplacian of the orbitals.  Section~\ref{sec:speed} compares the time cost in QMC calculations of polynomial methods and planewave expansions.  Section~\ref{sec:memory} compares the memory requirements of the polynomial methods and planewaves.  Section~\ref{sec:conclusions} concludes that higher accuracy and lower memory requirement make smoothing B-splines, with a separate approximation for the Laplacian, the best choice.  The appendix describes the details of the approximating polynomials.

\section{Methods} \label{sec:methods}

The four methods of approximating the planewave-represented single-particle orbitals with polynomials that we study in this paper are: interpolating Lagrange polynomials\cite{Lagrange}, interpolating piecewise-polynomial-form splines (pp-splines)\cite{deBoor01} (often simply called interpolating splines), and basis-form splines (B-splines) (both interpolating\cite{B_spline,deBoor01} and smoothing\cite{Hernandez97,Alfe04}).  For the pp-splines, we employ the Princeton \textsc{pspline} package\cite{PrincetonSpline}, and, for interpolating B-splines, the \textsc{einspline} package\cite{Einspline}, interfaced to the \textsc{champ} QMC program\cite{Cha-PROG-XX}.

{\it Common features.}  The methods share several aspects.  They construct the orbital approximation by a trivariate polynomial tensor product.  Each of the methods can employ polynomials of arbitrary order, $n$.  We use cubic polynomials, the customary choice.  The methods transform the cartesian coordinates to reduced coordinates prior to evaluating the polynomial approximation (see Eq.~(\ref{eq:reducedcoordinate}) and the Coordinate paragraph).  They use a grid of real-space points with associated coefficients and have a natural grid spacing defined by the highest-energy planewave in the planewave sum representing the orbital (see Eq.~(\ref{eq:natural_spacing}) and the Grid paragraph).  They share two possibilities for evaluating the required derivatives of a polynomial-represented function: (1) derivatives of the polynomials or (2) separate polynomial approximations of the planewave-represented derivatives.

{\it Distinctive features.}  As Lagrange-interpolated functions have discontinuous derivatives at the grid points, ensuring continuity in the derivative-dependent energies requires a separate interpolation for each of the components of the gradient and for the Laplacian of the orbitals, increasing the memory requirement by a factor of five.  In contrast to Lagrange interpolation, splines of degree $n$ have continuous derivatives up to order $n-1$ at the grid points, and, thus, the gradient and Laplacian of the splined function can approximate the gradient and Laplacian of the planewave sum, though this choice leads to a loss of accuracy. Spline functions have two free parameters in each dimension that are used to set the boundary conditions.  Since planewave-based orbitals are periodic, we choose the boundary conditions to have matching first and second derivatives at the boundaries in each dimension.  The formulation of both B-splines and pp-splines may be either interpolating (exact function values at the grid points)\cite{Einspline} or smoothing\cite{Hernandez97,Alfe04}. Smoothing splines are advantageous when the data is noisy, but this is not the case in our application.  Instead our rationale is the following: the planewave coefficients of each orbital specify that orbital, and the particular form of smoothing spline we use\cite{Hernandez97,Alfe04} is constructed to exactly reproduce the nonzero coefficients (see~\ref{smoothing_spline}).  Since fixing the values at the grid points and specifying the boundary conditions uniquely determines the interpolating spline function, interpolating B-splines and pp-splines yield identical function values\cite{deBoor01}. Due to the reduced number of coefficients stored per point, interpolating B-splines are preferable to pp-splines provided the time required for their evaluation is not larger than for pp-splines.

{\it Grid.}  Each of the interpolation methods permits either uniform or nonuniform grids.  For simplicity, we employ uniform grids, but the number of grid points in each dimension needs not be the same.  The highest-energy planewave in the planewave sum representing a given orbital defines a natural maximal grid spacing, above which short length scale information is lost.  One point per maximum and minimum of the highest-energy planewave $G_{\rm max}$, or two points per wavelength, $\lambda_{\rm min} = 2 \pi / G_{\rm max}$, defines this natural spacing $h_{\rm natural}$:
\begin{equation}
  h_{\rm natural} =
  \frac{\lambda_{\rm min}}{2} = \frac{\pi}{G_{\rm max}}.
  \label{eq:natural_spacing}
\end{equation}

{\it Coordinates.}  To simplify the form of the polynomials for the evaluation of the splines, we formulate the methods such that the point where the function is evaluated lies in the interval $[0,1)$ in each dimension. The spline evaluation for a given point requires
the coefficients at the four neighboring grid points in each dimension and requires
transforming the Cartesian coordinates to reduced coordinates. The
primitive-cell lattice vectors of the crystal ${\bf a}_i$ need not be orthogonal.
The reduced vector, ${\tilde{\bf r}} = (\tilde{r}_1,\tilde{r}_2,\tilde{r}_3) =
(\tilde{x},\tilde{y},\tilde{z})$, corresponding to the Cartesian vector
$\rvec$, is
\begin{eqnarray}
	\tilde{r}_i = \ngridsubi \left( \rcrystal_i -
	\lfloor \rcrystal_i \rfloor \right) - \\ \nonumber
      \lfloor \ngridsubi \left(
	\rcrystal_i - \lfloor \rcrystal_i \rfloor \right) \rfloor,
	\label{eq:reducedcoordinate}
\end{eqnarray}
where $\lfloor \, \rfloor$ is the floor function, which returns the integer part of a number and $\bf a$ is the $3\times 3$ matrix of lattice vectors ${\bf a}_i$.  Multiplying the Cartesian coordinate by ${\bf a}^{-1}$ transforms it to crystal coordinates. Subtracting the integer part of the crystal coordinates forces the coordinate inside the primitive cell, restricting the magnitude of the coordinate values to be between zero and one.  Multiplying the crystal coordinates by the number of grid points $N_{{\rm grid},i}$ along the $i^{th}$ lattice vector transforms the crystal coordinates into units of the grid point interval.  Subtracting the integer part of the interval-unit crystal coordinate (i.e., the index of the grid point smaller than and closest to $r_i$) yields the reduced coordinate ${\tilde{\bf r}}$, each component of which is in the interval $[0,1)$.  To obtain the Cartesian coordinate gradient and Laplacian starting from the reduced coordinate gradient and Hessian requires we use the chain rule, yielding ${\bf \nabla}\phi({\bf r}) = \ngridsubi ({\bf a}^{-1})^{\rm T} \tilde{\bf \nabla}\phi(\tilde{\bf r})$ and $\nabla^{2}\phi({\bf r}) = \ngridsubi^2 \sum_{i,j}(({\bf a}^{-1})({\bf a}^{-1})^{\rm T})_{ij} \partial^{2}\phi(\tilde{\bf r})/\partial \tilde{r}_i \partial \tilde{r}_j$. \ref{sec:explicitforms} gives the explicit forms of the approximating functions.

\section{Results}
\label{sec:results}

Three quantities compare the performance of Lagrange interpolation, pp-spline interpolation, B-spline interpolation and B-spline smoothing, within quantum Monte Carlo calculations of periodic systems: (i) the {\it accuracy} in reproducing the planewave orbital values, (ii) the {\it speedup} from the planewave-based calculation, and (iii) the computer {\it memory} required.

The results presented in this section are obtained for single-particle orbitals at the $\Gamma$ point, that are obtained with the LDA exchange-correlation functional in density functional theory and Troullier-Martins pseudopotentials\cite{TroullierMartins91}. \footnote{Mg -- valence configuration: $2s^22p^63d^{0.1}$, cutoff radii: $= 1.20, 1.50, 1.80 \:{\rm a.u.}$ respectively.  O -- valence configuration: $2s^22p^43d^0$, cutoff radii $= 1.0,1.0,1.0\:{\rm a.u.}$ respectively with Hamann's generalized state method\cite{Hamann89} for the d-channel.  Si -- valence configuration: $3s^23p^23d^0$, cutoff radii $= 2.25,2.25,2.25\:{\rm a.u.}$ respectively.  Al -- valence configuration: $3s^23p^13d^0$, cutoff radii $= 2.28,2.28,2.28\,{\rm a.u.}$ respectively with the generalized state method for the d-channel.}

\subsection{Accuracy} \label{sec:accuracy}

To understand the accuracy of the polynomial approximation to the planewave sum in the context of QMC, we compare the error in the orbital value, gradient of the orbital, Laplacian of the orbital, the total VMC energy, and the root-mean-square fluctuation in the VMC local energy relative to the corresponding quantity computed using the planewave sum.  Increasing the planewave cutoff tests the accuracy of the approximations as the planewave basis becomes more complete.
\footnote{Reducing the grid spacing for fixed planewave cutoff results in all quantities converging to their planewave values for the selected cutoff\cite{supplementary}}.

We find that the choice of $k$-point in our test does not affect the conclusions as similar results were obtained for various $k$-points ($\Gamma$, L, and X high-symmetry points in diamond Si) and different simulation cell sizes (2, 8, 16, and 32-atom cells in diamond Si).  However, significant differences in accuracy of the approximations occur for the three different materials tested, diamond Si, fcc Al and rock-salt MgO\cite{supplementary}. Since the approximation methods show the greatest differences from each other for the case of MgO in the rock-salt structure, we focus here on those results.

\begin{figure}
    \begin{center}
    \includegraphics[width=\columnwidth]{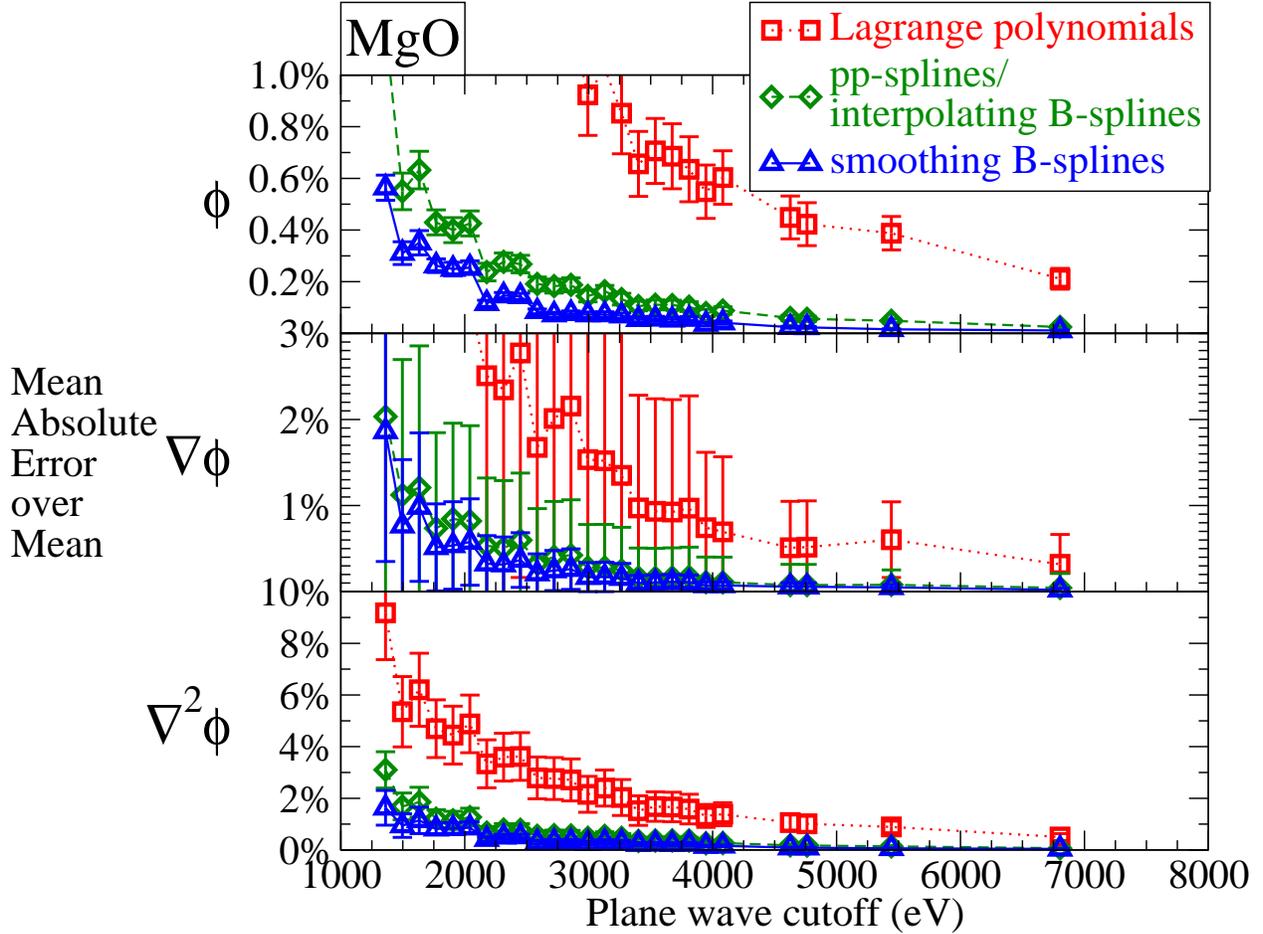}
     \end{center}
     \caption{(Color online) Relative mean absolute error of the
     orbitals, gradient of the orbitals, and Laplacian of the orbitals as a
     function of planewave cutoff in rock-salt MgO (chosen for greatest
     contrast in results) at natural grid spacing for each of the approximation
     methods:  Lagrange polynomials (squares), pp-splines and interpolating
     B-splines (diamonds), and smoothing B-splines (triangles).  Error bars indicate
     one standard deviation from the mean of the data point.  The error in the
     gradient and Laplacian of the orbitals is the error in a direct approximation
     of the gradient and Laplacian, respectively, of the orbital, not the gradient
     or Laplacian of an approximation of the orbital.  The large fluctuation in the
     error of the Laplacian is due to the small, stochastically chosen sample used
     to calculate the mean.  Smoothing B-splines show the smallest error relative to
     the planewave value at all grid spacings although the statistical uncertainty
     in the average over the components of the gradient obscures that result.
     } \label{fig:orbdorbddorb_error}
 \vskip 9mm
 \end{figure}
 
\begin{figure} \begin{center}
    \includegraphics[width=\columnwidth]{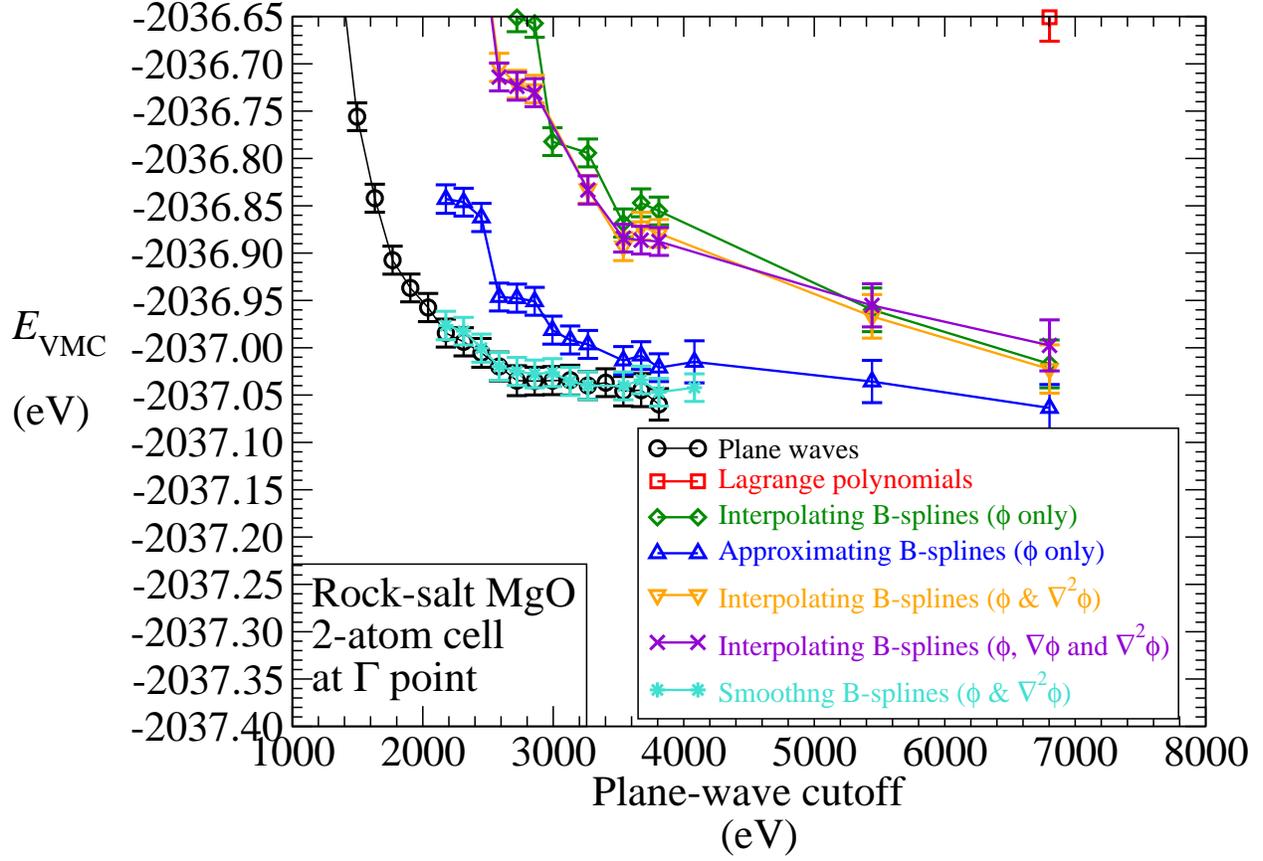}
    \end{center}
    \caption{(Color online) VMC total energy as a function of planewave basis
    cutoff in rock-salt MgO for each of the approximation methods at the
    natural grid spacing for that cutoff.  Error bars on points indicate one
    standard deviation of statistical uncertainty in the
    total energies.  Smoothing B-splines with separate approximations for
    the orbitals and the Laplacian of the orbitals lie within one standard
    deviation of statistical uncertainty of the planewave value for all
    cutoffs tested. (Note: Lagrange polynomials appear on the figure only at
    a cutoff of 250~Ha [$\approx 6800$ eV]).
} \label{fig:energy_with_cutoff}
\vskip 9mm
\end{figure}

\begin{figure}
  \begin{center}
    \includegraphics[width=\columnwidth]{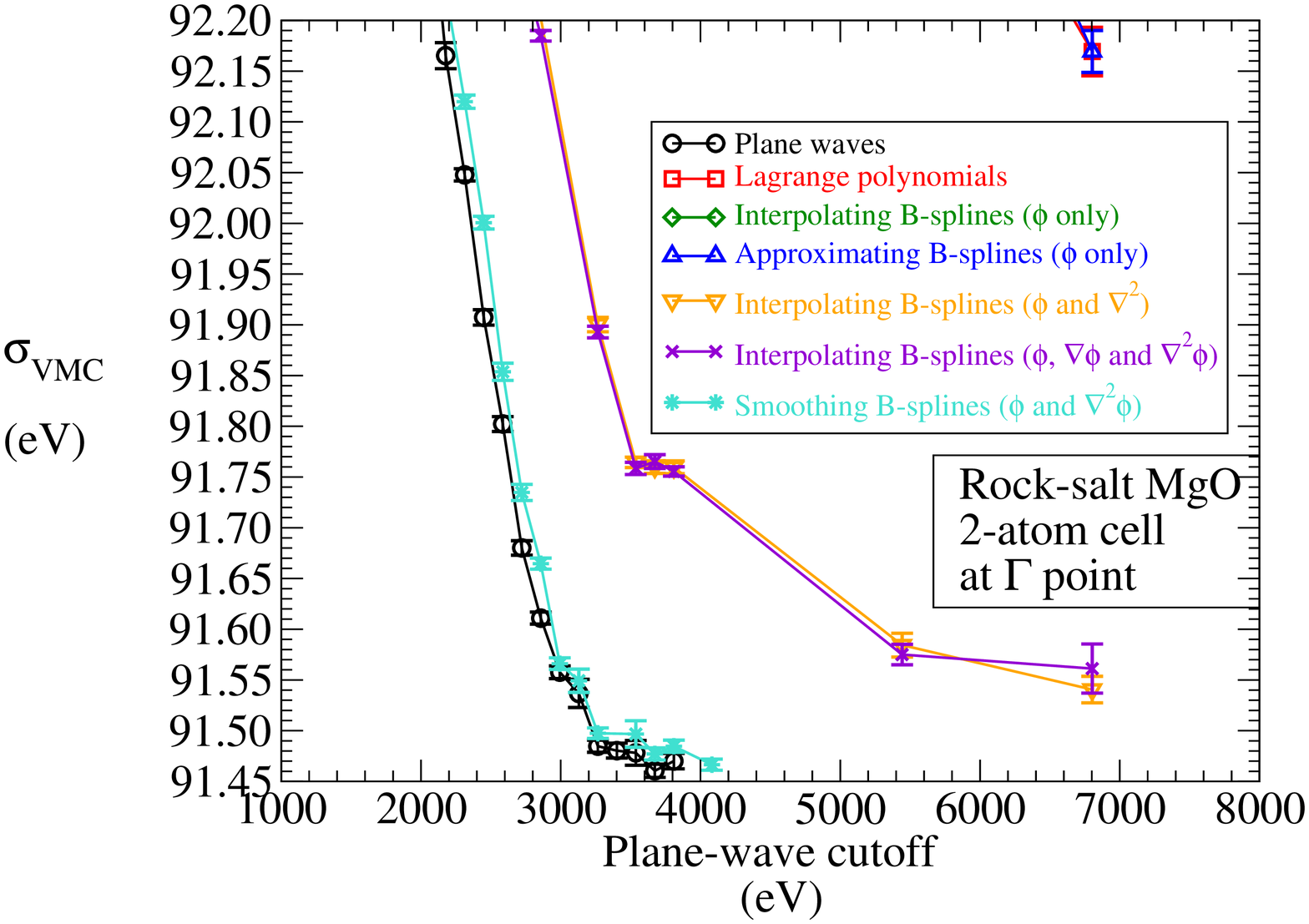}
  \end{center}
  \caption{(Color online) Root-mean-square fluctuation of the VMC local energy,
   $\sigma$, as a function of planewave basis cutoff in rock-salt MgO for each of
   the approximation methods at the natural grid spacing for that cutoff.  Error
   bars on points indicate one standard deviation of statistical uncertainty in
   the values.  Once the planewave value has converged, near 120~Ha ($\approx3270$
   eV), the smoothing B-splines with a separate approximation for the Laplacian is
   within statistical uncertainty of the planewave $\sigma$.  (Note: Lagrange
   polynomials and smoothing B-splines without separate approximation for the
   Laplacian appear on the figure only at a cutoff of 250~Ha[$\approx6800$
   eV]---interpolating B-splines without separate interpolation for the Laplacian
   do not appear at all on this scale)}
  \label{fig:sigma_with_cutoff}
\vskip 9mm
\end{figure}

Figure~\ref{fig:orbdorbddorb_error} shows the relative mean absolute error in the orbital, its gradient, and its Laplacian as a function of the planewave cutoff for the four approximation methods at natural grid spacing in rock-salt MgO.  The relative mean absolute error is the mean absolute error divided by the mean absolute value computed from the planewave sum.  Since pp-splines and interpolating B-splines give identical function values\cite{deBoor01}, they lie on a single curve and set of points. 

For each of the approximations, the gradient and Laplacian used by the QMC calculation can be obtained either by taking derivatives of the polynomial-approximated orbital, or, by constructing separate polynomial approximations for the gradient and the Laplacian of the planewave sum.  The central and lower panels of Figure~\ref{fig:orbdorbddorb_error} show the accuracy of separate approximations of the gradient and Laplacian of the planewave sum.  When separately approximating any derivatives of the orbitals, the resulting energy need not be an upper bound to the true energy, but the separate approximations recover the planewave value in the limit of infinite basis set.

Spline interpolation is more accurate than Lagrange interpolation for all planewave cutoffs.  Splines utilize all the tabulated function values (a global approximation) to enforce first and second derivative continuity across grid points, whereas Lagrange interpolation uses just the closest 64 points (a local approximation) and has derivative discontinuities at the grid points.  This leads to larger fluctuations in the error of Lagrange interpolation compared to splines.

Figures~\ref{fig:energy_with_cutoff} and~\ref{fig:sigma_with_cutoff} show the quantities of importance to QMC calculations, the total VMC energy $E_\mathrm{VMC}$ and the standard deviation of the local energy $\sigma_\mathrm{VMC}$, respectively, as a function of planewave cutoff in rock-salt MgO for the four approximations.  The deviations of $E_\mathrm{VMC}$ and $\sigma_\mathrm{VMC}$ from the planewave values reflect the errors in the orbitals, their gradients and Laplacian.  Smoothing B-splines are more accurate than interpolating splines, which in turn are more accurate than Lagrange interpolation.  Furthermore, separately approximating the Laplacian in the spline approximations significantly improves the accuracy of $E_\mathrm{VMC}$ and $\sigma_\mathrm{VMC}$.  The standard deviation of the local energy $\sigma_\mathrm{VMC}$ is more sensitive than the total energy $E_\mathrm{VMC}$ to the errors in the approximations because errors in the local energies partially cancel when averaging the local energy to obtain $E_\mathrm{VMC}$.
In all systems tested, convergence of $E_\mathrm{VMC}$ to within 1~mHa is observed for planewave cutoff energies 9-25~Ha smaller than for the convergence of $\sigma_\mathrm{VMC}$ to the same level.

\subsection{Speedup}
\label{sec:speed}
\begin{figure}
  \begin{center}
	\includegraphics[width=\columnwidth]{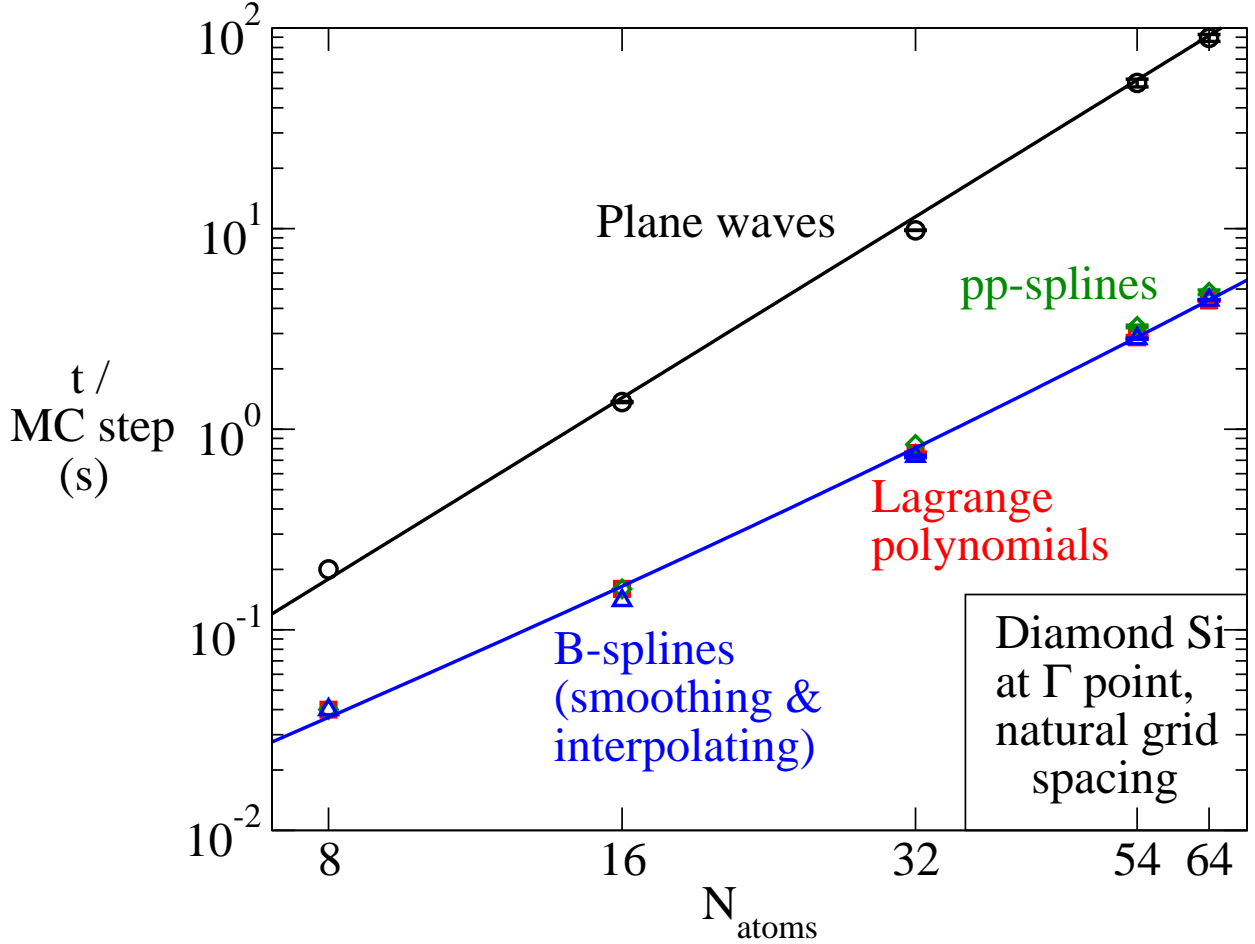}
  \end{center}

  \caption{(Color online) VMC time per Monte Carlo step versus number of atoms.
  All three approximations speed up the planewave calculation by a factor of
  $O(N)$ with nearly the same prefactor, recommending all approximation
  methods equally {\em on the basis of speed}.  
  Further optimization of the smoothing B-splines routines has reduced the time
  scaling prefactor by an order of magnitude from the data shown here.
  The similarity in form of the approximation methods
  (see Eqs.~(\ref{eq:lagrangeinterpolation3}),~(\ref{eq:pp-splineinterpolation3}),~(\ref{eq:b-splineapproximation}))
  and the fact that all methods need to retrieve a similar number of
  coefficients from memory
  account for the similarity in evaluation speed
  despite differences in approximation properties. 
  }
  \label{fig:speed}
\end{figure}

Figure~\ref{fig:speed} illustrates that the three methods of polynomial approximation speed up the planewave calculation by the same factor which scales as $O(N)$.  Tests on three different computer platforms (3.0 GHz Intel Pentium 4, 2.4 GHz Intel Xeon, 900 MHz Intel Itanium 2) show that the time scaling (in seconds) with the number of atoms $N$ for the approximating polynomials is of the order of $10^{-4}\,N^2 + 10^{-6}\,N^3$ compared to the scaling for planewaves of $10^{-3}\,N^3$.  The difference in computational time between the Lagrange polynomials and the B-splines is less than 10\% and varies between the different computers.

While pp-splines store eight coefficients at each grid point, and Lagrange interpolation and B-splines store just one, all methods require accessing the same number of coefficients (64 for cubic polynomials in 3D) from memory for each function evaluation.  However, Lagrange and B-splines access one coefficient from each of the 64 nearest-neighbor points whereas pp-splines access eight coefficients from each of the eight nearest-neighbor points.  This reduction in accessed neighbor points could make pp-splines faster since the data access is more local.  However, the calculations show that, for the implementation of pp-splines used here\cite{PrincetonSpline}, no speedup occurs in practice.  Additionally, further optimization of the smoothing B-splines routines has reduced the time scaling prefactor by an order of magnitude. 

\subsection{Memory}
\label{sec:memory}
At the natural grid spacing, the polynomial approximations store a total number of values equal to or greater than the number of planewaves.  \footnote{If the planewave expansion includes only planewaves that lie within a parallelepiped defined by three reciprocal lattice vectors, then the number of grid points at natural spacing (see Eq.~(\ref{eq:natural_spacing})) equals the number of planewave coefficients.  However, it is customary to include planewaves in a sphere up to some energy cutoff in the planewave sum, in which case the number of grid points is larger than the number of planewave coefficients.
For example, in a cubic lattice, the ratio of the number of grip points at natural
grid spacing to the number of planewaves is the ratio of the volume of a cube
to the volume of the inscribed sphere, namely $6/\pi$.
} Trivariate, cubic pp-splines store eight values per grid point for each function, namely the function values, the three second derivatives along each direction, three mixed fourth-order derivatives, and one mixed sixth-order derivative (see \citep{PrincetonSpline}] or Eq.~(\ref{eq:pp-splinecoefficients3}) for details).  Lagrange interpolation and B-splines store only one value per grid point for each function.  In the case of Lagrange interpolation, the stored values are the function values, whereas, for B-splines, the stored values are the derived B-spline coefficients (see \ref{sec:explicitforms}).

All the approximations can obtain the gradient and the Laplacian by either taking appropriate derivatives of the splined functions or by generating separate approximations for the gradient and the Laplacian.  Separate approximations for the Laplacian increase the memory requirement by a factor of two, and, separate approximations for the Laplacian and the gradient increase the memory requirement by a factor of five.  Since the gradient and Laplacian of the Lagrange interpolation are not continuous, we always use separate approximations for the Laplacian and the gradient when using Lagrange interpolation.  For splines, the increased accuracy achieved by using separate approximations warrants using separate approximations for the Laplacian but not for the gradient.

\section{Conclusions}
\label{sec:conclusions}
The four polynomial approximation methods -- interpolating Lagrange polynomials, interpolating pp-splines, interpolating B-splines, and smoothing B-splines -- speed up planewave-based quantum Monte Carlo (QMC) calculations by $O(N)$, where $N$ is the number of atoms in the system.  At natural grid spacing, smoothing B-splines are more accurate than interpolating splines, which are in turn more accurate than Lagrange interpolation for all planewave cutoff values tested.  Separately approximating the Laplacian of the orbitals results in the total energy and root-mean-square fluctuation of the local energy to be closest to the values obtained using the planewave sum.  High accuracy and low memory requirement make smoothing B-splines, with the Laplacian splined separately from the orbitals, the best choice for approximating planewave-based orbitals in QMC calculations.

\section{Acknowledgments}
This work was supported by the Department of Energy Basic Energy Sciences, Division of Materials Sciences (DE-FG02-99ER45795 and DE-FG05-08OR23339), and the National Science Foundation (CHE-1112097 and DMR-1056587).  Computational resources were provided by the Ohio Supercomputing Center, the National Center for Supercomputing Applications, and the National Energy Research Scientific Computing Center (supported by the Office of Science of the U.S. Department of Energy under Contract No. DE-AC02-05CH11231).  We thank Neil Drummond, Ken Esler, Jeongnim Kim, Mike Towler, and Andrew Williamson for helpful discussions, Ken Esler for recommending and helping with implementation of his Einspline library, and Jos{\'e} Lu{\'i}s Martins for the use of his pseudopotential generation and density functional programs.

\bibliographystyle{apsrev}
\bibliography{Bibliography}

\begin{thebibliography}{53}
\expandafter\ifx\csname natexlab\endcsname\relax\def\natexlab#1{#1}\fi
\expandafter\ifx\csname bibnamefont\endcsname\relax
  \def\bibnamefont#1{#1}\fi
\expandafter\ifx\csname bibfnamefont\endcsname\relax
  \def\bibfnamefont#1{#1}\fi
\expandafter\ifx\csname citenamefont\endcsname\relax
  \def\citenamefont#1{#1}\fi
\expandafter\ifx\csname url\endcsname\relax
  \def\url#1{\texttt{#1}}\fi
\expandafter\ifx\csname urlprefix\endcsname\relax\def\urlprefix{URL }\fi
\providecommand{\bibinfo}[2]{#2}
\providecommand{\eprint}[2][]{\url{#2}}

\bibitem[{\citenamefont{Foulkes et~al.}(2001)\citenamefont{Foulkes, Mitas,
  Needs, and Rajagopal}}]{QMCreview}
\bibinfo{author}{\bibfnamefont{W.~M.~C.} \bibnamefont{Foulkes}},
  \bibinfo{author}{\bibfnamefont{L.}~\bibnamefont{Mitas}},
  \bibinfo{author}{\bibfnamefont{R.~J.} \bibnamefont{Needs}}, \bibnamefont{and}
  \bibinfo{author}{\bibfnamefont{G.}~\bibnamefont{Rajagopal}},
  \bibinfo{journal}{Rev. Mod. Phys.} \textbf{\bibinfo{volume}{73}},
  \bibinfo{pages}{33} (\bibinfo{year}{2001}).

\bibitem[{\citenamefont{Nightingale and Umrigar}(1999)}]{NigUmr-BOOK-99}
\bibinfo{editor}{\bibfnamefont{M.~P.} \bibnamefont{Nightingale}}
  \bibnamefont{and} \bibinfo{editor}{\bibfnamefont{C.~J.}
  \bibnamefont{Umrigar}}, eds., \emph{\bibinfo{title}{Quantum Monte Carlo
  Methods in Physics and Chemistry}}, NATO ASI Ser. C 525
  (\bibinfo{publisher}{Kluwer}, \bibinfo{address}{Dordrecht},
  \bibinfo{year}{1999}).

\bibitem[{\citenamefont{Koloren\v{c} and Mitas}(2011)}]{Kolorenc11}
\bibinfo{author}{\bibfnamefont{J.}~\bibnamefont{Koloren\v{c}}}
  \bibnamefont{and} \bibinfo{author}{\bibfnamefont{L.}~\bibnamefont{Mitas}},
  \bibinfo{journal}{Reports on Progress in Physics}
  \textbf{\bibinfo{volume}{74}}, \bibinfo{pages}{026502}
  (\bibinfo{year}{2011}).

\bibitem[{\citenamefont{Anderson}(1975)}]{Anderson77}
\bibinfo{author}{\bibfnamefont{J.~B.} \bibnamefont{Anderson}},
  \bibinfo{journal}{J. Chem. Phys.} \textbf{\bibinfo{volume}{63}},
  \bibinfo{pages}{1499} (\bibinfo{year}{1975}).

\bibitem[{\citenamefont{Yao et~al.}(1996)\citenamefont{Yao, Xu, and
  Wang}}]{Yao96}
\bibinfo{author}{\bibfnamefont{G.}~\bibnamefont{Yao}},
  \bibinfo{author}{\bibfnamefont{J.~G.} \bibnamefont{Xu}}, \bibnamefont{and}
  \bibinfo{author}{\bibfnamefont{X.~W.} \bibnamefont{Wang}},
  \bibinfo{journal}{Phys. Rev. B} \textbf{\bibinfo{volume}{54}},
  \bibinfo{pages}{8393} (\bibinfo{year}{1996}).

\bibitem[{\citenamefont{Gaudoin et~al.}(2002)\citenamefont{Gaudoin, Foulkes,
  and Rajagopal}}]{Gaudoin02}
\bibinfo{author}{\bibfnamefont{R.}~\bibnamefont{Gaudoin}},
  \bibinfo{author}{\bibfnamefont{W.~M.~C.} \bibnamefont{Foulkes}},
  \bibnamefont{and}
  \bibinfo{author}{\bibfnamefont{G.}~\bibnamefont{Rajagopal}},
  \bibinfo{journal}{Journal of Physics: Condensed Matter}
  \textbf{\bibinfo{volume}{14}}, \bibinfo{pages}{8787} (\bibinfo{year}{2002}),
  \urlprefix\url{http://stacks.iop.org/0953-8984/14/8787}.

\bibitem[{\citenamefont{Hood et~al.}(2003)\citenamefont{Hood, Kent, Needs, and
  Briddon}}]{Hood03}
\bibinfo{author}{\bibfnamefont{R.~Q.} \bibnamefont{Hood}},
  \bibinfo{author}{\bibfnamefont{P.~R.~C.} \bibnamefont{Kent}},
  \bibinfo{author}{\bibfnamefont{R.~J.} \bibnamefont{Needs}}, \bibnamefont{and}
  \bibinfo{author}{\bibfnamefont{P.~R.} \bibnamefont{Briddon}},
  \bibinfo{journal}{Phys. Rev. Lett.} \textbf{\bibinfo{volume}{91}},
  \bibinfo{pages}{076403} (\bibinfo{year}{2003}).

\bibitem[{\citenamefont{Maezono et~al.}(2003)\citenamefont{Maezono, Towler,
  Lee, and Needs}}]{Maezono03}
\bibinfo{author}{\bibfnamefont{R.}~\bibnamefont{Maezono}},
  \bibinfo{author}{\bibfnamefont{M.~D.} \bibnamefont{Towler}},
  \bibinfo{author}{\bibfnamefont{Y.}~\bibnamefont{Lee}}, \bibnamefont{and}
  \bibinfo{author}{\bibfnamefont{R.~J.} \bibnamefont{Needs}},
  \bibinfo{journal}{Phys. Rev. B} \textbf{\bibinfo{volume}{68}},
  \bibinfo{pages}{165103} (\bibinfo{year}{2003}).

\bibitem[{\citenamefont{Needs and Towler}(2003)}]{Needs03}
\bibinfo{author}{\bibfnamefont{R.~J.} \bibnamefont{Needs}} \bibnamefont{and}
  \bibinfo{author}{\bibfnamefont{M.~D.} \bibnamefont{Towler}},
  \bibinfo{journal}{International Journal of Modern Physics B}
  \textbf{\bibinfo{volume}{17}}, \bibinfo{pages}{5425 } (\bibinfo{year}{2003}).

\bibitem[{\citenamefont{Alf\`e et~al.}(2004)\citenamefont{Alf\`e, Gillan,
  Towler, and Needs}}]{Alfe04b}
\bibinfo{author}{\bibfnamefont{D.}~\bibnamefont{Alf\`e}},
  \bibinfo{author}{\bibfnamefont{M.~J.} \bibnamefont{Gillan}},
  \bibinfo{author}{\bibfnamefont{M.~D.} \bibnamefont{Towler}},
  \bibnamefont{and} \bibinfo{author}{\bibfnamefont{R.~J.} \bibnamefont{Needs}},
  \bibinfo{journal}{Phys. Rev. B} \textbf{\bibinfo{volume}{70}},
  \bibinfo{pages}{214102} (\bibinfo{year}{2004}).

\bibitem[{\citenamefont{Alf\`e and Gillan}(2005)}]{Alfe05a}
\bibinfo{author}{\bibfnamefont{D.}~\bibnamefont{Alf\`e}} \bibnamefont{and}
  \bibinfo{author}{\bibfnamefont{M.~J.} \bibnamefont{Gillan}},
  \bibinfo{journal}{Phys. Rev. B} \textbf{\bibinfo{volume}{71}},
  \bibinfo{pages}{220101} (\bibinfo{year}{2005}).

\bibitem[{\citenamefont{Alf\`e et~al.}(2005)\citenamefont{Alf\`e, Alfredsson,
  Brodholt, Gillan, Towler, and Needs}}]{Alfe05b}
\bibinfo{author}{\bibfnamefont{D.}~\bibnamefont{Alf\`e}},
  \bibinfo{author}{\bibfnamefont{M.}~\bibnamefont{Alfredsson}},
  \bibinfo{author}{\bibfnamefont{J.}~\bibnamefont{Brodholt}},
  \bibinfo{author}{\bibfnamefont{M.~J.} \bibnamefont{Gillan}},
  \bibinfo{author}{\bibfnamefont{M.~D.} \bibnamefont{Towler}},
  \bibnamefont{and} \bibinfo{author}{\bibfnamefont{R.~J.} \bibnamefont{Needs}},
  \bibinfo{journal}{Phys. Rev. B} \textbf{\bibinfo{volume}{72}},
  \bibinfo{pages}{014114} (\bibinfo{year}{2005}).

\bibitem[{\citenamefont{Drummond and Needs}(2006)}]{Drummond06}
\bibinfo{author}{\bibfnamefont{N.~D.} \bibnamefont{Drummond}} \bibnamefont{and}
  \bibinfo{author}{\bibfnamefont{R.~J.} \bibnamefont{Needs}},
  \bibinfo{journal}{Phys. Rev. B} \textbf{\bibinfo{volume}{73}},
  \bibinfo{eid}{024107} (pages~\bibinfo{numpages}{8}) (\bibinfo{year}{2006}),
  \urlprefix\url{http://link.aps.org/abstract/PRB/v73/e024107}.

\bibitem[{\citenamefont{Batista et~al.}(2006)\citenamefont{Batista, Heyd,
  Hennig, Uberuaga, Martin, Scuseria, Umrigar, and Wilkins}}]{Batista06}
\bibinfo{author}{\bibfnamefont{E.~R.} \bibnamefont{Batista}},
  \bibinfo{author}{\bibfnamefont{J.}~\bibnamefont{Heyd}},
  \bibinfo{author}{\bibfnamefont{R.~G.} \bibnamefont{Hennig}},
  \bibinfo{author}{\bibfnamefont{B.~P.} \bibnamefont{Uberuaga}},
  \bibinfo{author}{\bibfnamefont{R.~L.} \bibnamefont{Martin}},
  \bibinfo{author}{\bibfnamefont{G.~E.} \bibnamefont{Scuseria}},
  \bibinfo{author}{\bibfnamefont{C.~J.} \bibnamefont{Umrigar}},
  \bibnamefont{and} \bibinfo{author}{\bibfnamefont{J.~W.}
  \bibnamefont{Wilkins}}, \bibinfo{journal}{Phys. Rev. B}
  \textbf{\bibinfo{volume}{74}}, \bibinfo{eid}{121102}
  (pages~\bibinfo{numpages}{4}) (\bibinfo{year}{2006}),
  \urlprefix\url{http://link.aps.org/abstract/PRB/v74/e121102}.

\bibitem[{\citenamefont{Maezono et~al.}(2007)\citenamefont{Maezono, Ma, Towler,
  and Needs}}]{Maezono07}
\bibinfo{author}{\bibfnamefont{R.}~\bibnamefont{Maezono}},
  \bibinfo{author}{\bibfnamefont{A.}~\bibnamefont{Ma}},
  \bibinfo{author}{\bibfnamefont{M.~D.} \bibnamefont{Towler}},
  \bibnamefont{and} \bibinfo{author}{\bibfnamefont{R.~J.} \bibnamefont{Needs}},
  \bibinfo{journal}{Phys. Rev. Lett.} \textbf{\bibinfo{volume}{98}},
  \bibinfo{eid}{025701} (pages~\bibinfo{numpages}{4}) (\bibinfo{year}{2007}),
  \urlprefix\url{http://link.aps.org/abstract/PRL/v98/e025701}.

\bibitem[{\citenamefont{Pozzo and Alf\`{e}}(2008)}]{Pozzo08}
\bibinfo{author}{\bibfnamefont{M.}~\bibnamefont{Pozzo}} \bibnamefont{and}
  \bibinfo{author}{\bibfnamefont{D.}~\bibnamefont{Alf\`{e}}},
  \bibinfo{journal}{Phys. Rev. B} \textbf{\bibinfo{volume}{77}},
  \bibinfo{eid}{104103} (pages~\bibinfo{numpages}{8}) (\bibinfo{year}{2008}),
  \urlprefix\url{http://link.aps.org/abstract/PRB/v77/e104103}.

\bibitem[{\citenamefont{Koloren\v{c} and Mitas}(2008)}]{Kolorenc08}
\bibinfo{author}{\bibfnamefont{J.}~\bibnamefont{Koloren\v{c}}}
  \bibnamefont{and} \bibinfo{author}{\bibfnamefont{L.}~\bibnamefont{Mitas}},
  \bibinfo{journal}{Phys. Rev. Lett.} \textbf{\bibinfo{volume}{101}},
  \bibinfo{eid}{185502} (pages~\bibinfo{numpages}{4}) (\bibinfo{year}{2008}),
  \urlprefix\url{http://link.aps.org/abstract/PRL/v101/e185502}.

\bibitem[{\citenamefont{Sola et~al.}(2009)\citenamefont{Sola, Brodholt, and
  Alf\`{e}}}]{Sola09}
\bibinfo{author}{\bibfnamefont{E.}~\bibnamefont{Sola}},
  \bibinfo{author}{\bibfnamefont{J.~P.} \bibnamefont{Brodholt}},
  \bibnamefont{and} \bibinfo{author}{\bibfnamefont{D.}~\bibnamefont{Alf\`{e}}},
  \bibinfo{journal}{Phys. Rev. B} \textbf{\bibinfo{volume}{79}},
  \bibinfo{eid}{024107} (pages~\bibinfo{numpages}{6}) (\bibinfo{year}{2009}),
  \urlprefix\url{http://link.aps.org/abstract/PRB/v79/e024107}.

\bibitem[{\citenamefont{Hennig et~al.}(2010)\citenamefont{Hennig, Wadehra,
  Driver, Parker, Umrigar, and Wilkins}}]{Hennig10}
\bibinfo{author}{\bibfnamefont{R.~G.} \bibnamefont{Hennig}},
  \bibinfo{author}{\bibfnamefont{A.}~\bibnamefont{Wadehra}},
  \bibinfo{author}{\bibfnamefont{K.~P.} \bibnamefont{Driver}},
  \bibinfo{author}{\bibfnamefont{W.~D.} \bibnamefont{Parker}},
  \bibinfo{author}{\bibfnamefont{C.~J.} \bibnamefont{Umrigar}},
  \bibnamefont{and} \bibinfo{author}{\bibfnamefont{J.~W.}
  \bibnamefont{Wilkins}}, \bibinfo{journal}{Phys. Rev. B}
  \textbf{\bibinfo{volume}{82}}, \bibinfo{pages}{014101}
  (\bibinfo{year}{2010}).

\bibitem[{\citenamefont{Driver et~al.}(2010)\citenamefont{Driver, Cohen, Wu,
  Militzer, Ríos, Towler, Needs, and Wilkins}}]{Driver10}
\bibinfo{author}{\bibfnamefont{K.~P.} \bibnamefont{Driver}},
  \bibinfo{author}{\bibfnamefont{R.~E.} \bibnamefont{Cohen}},
  \bibinfo{author}{\bibfnamefont{Z.}~\bibnamefont{Wu}},
  \bibinfo{author}{\bibfnamefont{B.}~\bibnamefont{Militzer}},
  \bibinfo{author}{\bibfnamefont{P.~L.} \bibnamefont{Ríos}},
  \bibinfo{author}{\bibfnamefont{M.~D.} \bibnamefont{Towler}},
  \bibinfo{author}{\bibfnamefont{R.~J.} \bibnamefont{Needs}}, \bibnamefont{and}
  \bibinfo{author}{\bibfnamefont{J.~W.} \bibnamefont{Wilkins}},
  \bibinfo{journal}{Proceedings of the National Academy of Sciences}
  \textbf{\bibinfo{volume}{107}}, \bibinfo{pages}{9519} (\bibinfo{year}{2010}).

\bibitem[{\citenamefont{Maezono et~al.}(2010)\citenamefont{Maezono, Drummond,
  Ma, and Needs}}]{Maezono10}
\bibinfo{author}{\bibfnamefont{R.}~\bibnamefont{Maezono}},
  \bibinfo{author}{\bibfnamefont{N.~D.} \bibnamefont{Drummond}},
  \bibinfo{author}{\bibfnamefont{A.}~\bibnamefont{Ma}}, \bibnamefont{and}
  \bibinfo{author}{\bibfnamefont{R.~J.} \bibnamefont{Needs}},
  \bibinfo{journal}{Phys. Rev. B} \textbf{\bibinfo{volume}{82}},
  \bibinfo{pages}{184108} (\bibinfo{year}{2010}).

\bibitem[{\citenamefont{Parker et~al.}(2011)\citenamefont{Parker, Wilkins, and
  Hennig}}]{Parker11}
\bibinfo{author}{\bibfnamefont{W.~D.} \bibnamefont{Parker}},
  \bibinfo{author}{\bibfnamefont{J.~W.} \bibnamefont{Wilkins}},
  \bibnamefont{and} \bibinfo{author}{\bibfnamefont{R.~G.}
  \bibnamefont{Hennig}}, \bibinfo{journal}{physica status solidi (b)}
  \textbf{\bibinfo{volume}{248}}, \bibinfo{pages}{267} (\bibinfo{year}{2011}),
  ISSN \bibinfo{issn}{1521-3951}.

\bibitem[{\citenamefont{Abbasnejad et~al.}(2012)\citenamefont{Abbasnejad,
  Shojaee, Mohammadizadeh, Alaei, and Maezono}}]{Abbasnejad12}
\bibinfo{author}{\bibfnamefont{M.}~\bibnamefont{Abbasnejad}},
  \bibinfo{author}{\bibfnamefont{E.}~\bibnamefont{Shojaee}},
  \bibinfo{author}{\bibfnamefont{M.~R.} \bibnamefont{Mohammadizadeh}},
  \bibinfo{author}{\bibfnamefont{M.}~\bibnamefont{Alaei}}, \bibnamefont{and}
  \bibinfo{author}{\bibfnamefont{R.}~\bibnamefont{Maezono}},
  \bibinfo{journal}{Applied Physics Letters} \textbf{\bibinfo{volume}{100}},
  \bibinfo{eid}{261902} (\bibinfo{year}{2012}).

\bibitem[{\citenamefont{Schwarz et~al.}(2012)\citenamefont{Schwarz,
  Sundararaman, Letchworth-Weaver, Arias, and Hennig}}]{Schwarz12}
\bibinfo{author}{\bibfnamefont{K.~A.} \bibnamefont{Schwarz}},
  \bibinfo{author}{\bibfnamefont{R.}~\bibnamefont{Sundararaman}},
  \bibinfo{author}{\bibfnamefont{K.}~\bibnamefont{Letchworth-Weaver}},
  \bibinfo{author}{\bibfnamefont{T.~A.} \bibnamefont{Arias}}, \bibnamefont{and}
  \bibinfo{author}{\bibfnamefont{R.~G.} \bibnamefont{Hennig}},
  \bibinfo{journal}{Phys. Rev. B} \textbf{\bibinfo{volume}{85}},
  \bibinfo{pages}{201102} (\bibinfo{year}{2012}),
  \urlprefix\url{http://link.aps.org/doi/10.1103/PhysRevB.85.201102}.

\bibitem[{\citenamefont{Hood et~al.}(2012)\citenamefont{Hood, Kent, and
  Reboredo}}]{Hood12}
\bibinfo{author}{\bibfnamefont{R.~Q.} \bibnamefont{Hood}},
  \bibinfo{author}{\bibfnamefont{P.~R.~C.} \bibnamefont{Kent}},
  \bibnamefont{and} \bibinfo{author}{\bibfnamefont{F.~A.}
  \bibnamefont{Reboredo}}, \bibinfo{journal}{Phys. Rev. B}
  \textbf{\bibinfo{volume}{85}}, \bibinfo{pages}{134109}
  (\bibinfo{year}{2012}).

\bibitem[{\citenamefont{Azadi et~al.}(2013)\citenamefont{Azadi, Foulkes, and
  Kühne}}]{Azadi13}
\bibinfo{author}{\bibfnamefont{S.}~\bibnamefont{Azadi}},
  \bibinfo{author}{\bibfnamefont{W.~M.~C.} \bibnamefont{Foulkes}},
  \bibnamefont{and} \bibinfo{author}{\bibfnamefont{T.~D.}
  \bibnamefont{Kühne}}, \bibinfo{journal}{New Journal of Physics}
  \textbf{\bibinfo{volume}{15}}, \bibinfo{pages}{113005}
  (\bibinfo{year}{2013}).

\bibitem[{\citenamefont{Ertekin et~al.}(2013)\citenamefont{Ertekin, Wagner, and
  Grossman}}]{Ertekin13}
\bibinfo{author}{\bibfnamefont{E.}~\bibnamefont{Ertekin}},
  \bibinfo{author}{\bibfnamefont{L.~K.} \bibnamefont{Wagner}},
  \bibnamefont{and} \bibinfo{author}{\bibfnamefont{J.~C.}
  \bibnamefont{Grossman}}, \bibinfo{journal}{Phys. Rev. B}
  \textbf{\bibinfo{volume}{87}}, \bibinfo{pages}{155210}
  (\bibinfo{year}{2013}).

\bibitem[{\citenamefont{Shulenburger and Mattsson}(2013)}]{Shulenburger13}
\bibinfo{author}{\bibfnamefont{L.}~\bibnamefont{Shulenburger}}
  \bibnamefont{and} \bibinfo{author}{\bibfnamefont{T.~R.}
  \bibnamefont{Mattsson}}, \bibinfo{journal}{Phys. Rev. B}
  \textbf{\bibinfo{volume}{88}}, \bibinfo{pages}{245117}
  (\bibinfo{year}{2013}).

\bibitem[{\citenamefont{Chen et~al.}(2014)\citenamefont{Chen, Ren, Li,
  Alf\`{e}, and Wang}}]{Chen14}
\bibinfo{author}{\bibfnamefont{J.}~\bibnamefont{Chen}},
  \bibinfo{author}{\bibfnamefont{X.}~\bibnamefont{Ren}},
  \bibinfo{author}{\bibfnamefont{X.-Z.} \bibnamefont{Li}},
  \bibinfo{author}{\bibfnamefont{D.}~\bibnamefont{Alf\`{e}}}, \bibnamefont{and}
  \bibinfo{author}{\bibfnamefont{E.}~\bibnamefont{Wang}}, \bibinfo{journal}{The
  Journal of Chemical Physics} \textbf{\bibinfo{volume}{141}},
  \bibinfo{eid}{024501} (\bibinfo{year}{2014}).

\bibitem[{\citenamefont{Benali et~al.}(2014)\citenamefont{Benali, Shulenburger,
  Romero, Kim, and von Lilienfeld}}]{Benali14}
\bibinfo{author}{\bibfnamefont{A.}~\bibnamefont{Benali}},
  \bibinfo{author}{\bibfnamefont{L.}~\bibnamefont{Shulenburger}},
  \bibinfo{author}{\bibfnamefont{N.~A.} \bibnamefont{Romero}},
  \bibinfo{author}{\bibfnamefont{J.}~\bibnamefont{Kim}}, \bibnamefont{and}
  \bibinfo{author}{\bibfnamefont{O.~A.} \bibnamefont{von Lilienfeld}},
  \bibinfo{journal}{Journal of Chemical Theory and Computation}
  \textbf{\bibinfo{volume}{10}}, \bibinfo{pages}{3417} (\bibinfo{year}{2014}).

\bibitem[{\citenamefont{Azadi et~al.}(2014)\citenamefont{Azadi, Monserrat,
  Foulkes, and Needs}}]{Azadi14}
\bibinfo{author}{\bibfnamefont{S.}~\bibnamefont{Azadi}},
  \bibinfo{author}{\bibfnamefont{B.}~\bibnamefont{Monserrat}},
  \bibinfo{author}{\bibfnamefont{W.~M.~C.} \bibnamefont{Foulkes}},
  \bibnamefont{and} \bibinfo{author}{\bibfnamefont{R.~J.} \bibnamefont{Needs}},
  \bibinfo{journal}{Phys. Rev. Lett.} \textbf{\bibinfo{volume}{112}},
  \bibinfo{pages}{165501} (\bibinfo{year}{2014}).

\bibitem[{\citenamefont{Foyevtsova et~al.}(2014)\citenamefont{Foyevtsova,
  Krogel, Kim, Kent, Dagotto, and Reboredo}}]{Foyevtsova14}
\bibinfo{author}{\bibfnamefont{K.}~\bibnamefont{Foyevtsova}},
  \bibinfo{author}{\bibfnamefont{J.~T.} \bibnamefont{Krogel}},
  \bibinfo{author}{\bibfnamefont{J.}~\bibnamefont{Kim}},
  \bibinfo{author}{\bibfnamefont{P.~R.~C.} \bibnamefont{Kent}},
  \bibinfo{author}{\bibfnamefont{E.}~\bibnamefont{Dagotto}}, \bibnamefont{and}
  \bibinfo{author}{\bibfnamefont{F.~A.} \bibnamefont{Reboredo}},
  \bibinfo{journal}{Phys. Rev. X} \textbf{\bibinfo{volume}{4}},
  \bibinfo{pages}{031003} (\bibinfo{year}{2014}).

\bibitem[{\citenamefont{Jastrow}(1955)}]{JastrowFactor}
\bibinfo{author}{\bibfnamefont{R.}~\bibnamefont{Jastrow}},
  \bibinfo{journal}{Phys. Rev.} \textbf{\bibinfo{volume}{98}},
  \bibinfo{pages}{1479} (\bibinfo{year}{1955}).

\bibitem[{\citenamefont{Slater}(1932)}]{SlaterDeterminant}
\bibinfo{author}{\bibfnamefont{J.~C.} \bibnamefont{Slater}},
  \bibinfo{journal}{Phys. Rev.} \textbf{\bibinfo{volume}{42}},
  \bibinfo{pages}{33 } (\bibinfo{year}{1932}).

\bibitem[{\citenamefont{Williamson et~al.}(2001)\citenamefont{Williamson, Hood,
  and Grossman}}]{Williamson01}
\bibinfo{author}{\bibfnamefont{A.}~\bibnamefont{Williamson}},
  \bibinfo{author}{\bibfnamefont{R.~Q.} \bibnamefont{Hood}}, \bibnamefont{and}
  \bibinfo{author}{\bibfnamefont{J.~C.} \bibnamefont{Grossman}},
  \bibinfo{journal}{Phys. Rev. Lett.} \textbf{\bibinfo{volume}{87}},
  \bibinfo{pages}{246406} (\bibinfo{year}{2001}).

\bibitem[{\citenamefont{Alf\`e and Gillan}(2004)}]{Alfe04}
\bibinfo{author}{\bibfnamefont{D.}~\bibnamefont{Alf\`e}} \bibnamefont{and}
  \bibinfo{author}{\bibfnamefont{M.~J.} \bibnamefont{Gillan}},
  \bibinfo{journal}{Phys. Rev. B} \textbf{\bibinfo{volume}{70}},
  \bibinfo{pages}{161101(R)} (\bibinfo{year}{2004}), \bibinfo{note}{note that
  $\gamma_k$ should be on the other side of Eq.~6.}

\bibitem[{\citenamefont{Esler}()}]{Esler_unpub}
\bibinfo{author}{\bibfnamefont{K.}~\bibnamefont{Esler}},
  \bibinfo{note}{(unpublished)}.

\bibitem[{\citenamefont{Hern\'andez et~al.}(1997)\citenamefont{Hern\'andez,
  Gillan, and Goringe}}]{Hernandez97}
\bibinfo{author}{\bibfnamefont{E.}~\bibnamefont{Hern\'andez}},
  \bibinfo{author}{\bibfnamefont{M.~J.} \bibnamefont{Gillan}},
  \bibnamefont{and} \bibinfo{author}{\bibfnamefont{C.~M.}
  \bibnamefont{Goringe}}, \bibinfo{journal}{Phys. Rev. B}
  \textbf{\bibinfo{volume}{55}}, \bibinfo{pages}{13485} (\bibinfo{year}{1997}).

\bibitem[{Lag()}]{Lagrange}
\bibinfo{note}{Lagrange polynomial interpolation is a generalization of the
  earliest interpolation efforts, such as those by
  Newton.~\cite{PrincipiaInterpolation} Waring~\cite{Waring79} first published
  the general Lagrange form in 1779, but the method later acquired Lagrange's
  name because of his popular
  lectures~\cite{ElementaryMathematicsInterpolation} on the
  subject.~\cite{PearsonHistoryofInterpolation}}.

\bibitem[{\citenamefont{de~Boor}(2001)}]{deBoor01}
\bibinfo{author}{\bibfnamefont{C.}~\bibnamefont{de~Boor}},
  \emph{\bibinfo{title}{A Practical Guide to Splines}}
  (\bibinfo{publisher}{Springer}, \bibinfo{year}{2001}),
  \bibinfo{note}{{C}hapter IV defines local approximation and introduces
  pp-form splines. Chapter IX introduces B-form splines. Page 99 states
  B-splines can represent any piecewise-polynomial function, and page 101
  discusses the conversion between pp-form and B-form.}

\bibitem[{B_s()}]{B_spline}
\bibinfo{note}{The B-spline was so-named by Schoenberg~\cite{Curry47}}.

\bibitem[{Pri()}]{PrincetonSpline}
\urlprefix\url{http://w3.pppl.gov/ntcc/PSPLINE/}.

\bibitem[{Ein()}]{Einspline}
\urlprefix\url{http://einspline.sourceforge.net/}.

\bibitem[{Cha()}]{Cha-PROG-XX}
\bibinfo{note}{CHAMP, a quantum Monte Carlo program written by C. J. Umrigar,
  C. Filippi, Julien Toulouse and collaborators},
  \urlprefix\url{http://www.ccmr.cornell.edu/${\sim}$cyrus/champ.html}.

\bibitem[{\citenamefont{{N. Troullier and J. L.
  Martins}}(1991)}]{TroullierMartins91}
\bibinfo{author}{\bibnamefont{{N. Troullier and J. L. Martins}}},
  \bibinfo{journal}{Phys. Rev. B} \textbf{\bibinfo{volume}{43}},
  \bibinfo{pages}{1993} (\bibinfo{year}{1991}).

\bibitem[{sup()}]{supplementary}
\bibinfo{note}{See online supplementary data for tests on diamond Si and fcc Al
  and additional data for rock-salt MgO.}

\bibitem[{\citenamefont{Newton}(1848)}]{PrincipiaInterpolation}
\bibinfo{author}{\bibfnamefont{I.}~\bibnamefont{Newton}},
  \emph{\bibinfo{title}{Mathematical Principles of Natural Philosophy}}
  (\bibinfo{publisher}{Daniel Adee}, \bibinfo{year}{1848}),
  \bibinfo{note}{{B}ook 3, Lemma V}.

\bibitem[{\citenamefont{Waring}(1779)}]{Waring79}
\bibinfo{author}{\bibfnamefont{E.}~\bibnamefont{Waring}},
  \bibinfo{journal}{Philisophical Transactions of the Royal Society of London}
  \textbf{\bibinfo{volume}{69}}, \bibinfo{pages}{59} (\bibinfo{year}{1779}).

\bibitem[{\citenamefont{Lagrange}(1901)}]{ElementaryMathematicsInterpolation}
\bibinfo{author}{\bibfnamefont{J.~L.} \bibnamefont{Lagrange}},
  \emph{\bibinfo{title}{Lectures on Elementary Mathematics}}
  (\bibinfo{publisher}{The Open Court Publishing Company},
  \bibinfo{year}{1901}), \bibinfo{note}{{L}ecture V}.

\bibitem[{\citenamefont{Pearson}(1920)}]{PearsonHistoryofInterpolation}
\bibinfo{editor}{\bibfnamefont{K.}~\bibnamefont{Pearson}}, ed.,
  \emph{\bibinfo{title}{Tracts for Computers}} (\bibinfo{publisher}{University
  Press}, \bibinfo{year}{1920}), vol.~\bibinfo{volume}{2}, pp.
  \bibinfo{pages}{62--64}.

\bibitem[{\citenamefont{Curry and Schoenberg}(1947)}]{Curry47}
\bibinfo{author}{\bibfnamefont{H.~B.} \bibnamefont{Curry}} \bibnamefont{and}
  \bibinfo{author}{\bibfnamefont{I.~J.} \bibnamefont{Schoenberg}},
  \bibinfo{journal}{Bull. Amer. Math. Soc.} \textbf{\bibinfo{volume}{53}},
  \bibinfo{pages}{1114} (\bibinfo{year}{1947}).

\bibitem[{\citenamefont{Hamann}(1989)}]{Hamann89}
\bibinfo{author}{\bibfnamefont{D.~R.} \bibnamefont{Hamann}},
  \bibinfo{journal}{Phys. Rev. B} \textbf{\bibinfo{volume}{40}},
  \bibinfo{pages}{2980} (\bibinfo{year}{1989}).

\bibitem[{\citenamefont{Nord}(1967)}]{Nord67}
\bibinfo{author}{\bibfnamefont{S.}~\bibnamefont{Nord}}, \bibinfo{journal}{BIT
  Numerical Mathematics} \textbf{\bibinfo{volume}{7}}, \bibinfo{pages}{132}
  (\bibinfo{year}{1967}).

\end{thebibliography}

\appendix
\section{Explicit Forms of Approximation Methods}
\label{sec:explicitforms}
\subsection{Lagrange interpolation}
In one dimension, the Lagrange interpolation formula for a function $f(x)$ is
\beq
L(x) &=& \sum_{i=m-\lfloor \frac{n}{2} \rfloor}^{m+\lfloor \frac{n+1}{2} \rfloor} \ell_i(x) f(x_i)
\eeq
where the basis polynomials of order $n$ are
\begin{equation}
\label{eq:lagrangebasisgeneral}
\ell_i(x) = \prod_{j=m-\lfloor \frac{n}{2} \rfloor,j\neq i}^{m+\lfloor \frac{n+1}{2} \rfloor}\frac{x-x_{j}}{x_{i}-x_{j}}.
\end{equation}
where the grid point $m$ is such that $x_m \le x < x_{m+1}$.

A tensor product of the one-dimensional basis constructs basis functions for representing multidimensional functions.  Hence, the cubic Lagrange
interpolation formula for the 3-dimensional orbital using the reduced coordinates of Eq.~(\ref{eq:reducedcoordinate}) is
\begin{eqnarray}
\label{eq:lagrangeinterpolation3}
\lefteqn{\phi_{\rm Lagr}(\xt,\yt,\zt) =  }\\ \nonumber
   && \sum_{i=-1}^{2} \ell_i(\tilde{x}) \sum_{j=-1}^{2} \ell_j(\tilde{y}) \sum_{k=-1}^{2} \ell_k(\tilde{z}) \; \phi_{\rm PW}(i,j,k),
\end{eqnarray}
where
\begin{equation}
\label{eq:lagrangebasisgeneral3}
\ell_i(\xt) = \prod_{j=-1,j\neq i}^{2}\frac{\xt-j}{i-j}.
\end{equation}
or
\begin{subequations}
\label{eq:lagrangebasis}
\begin{align}
	\ell_{-1}(\tilde{x}) &= -\frac{1}{6} \tilde{x}(\tilde{x}-1)(\tilde{x}-2) \\ \nonumber
        & = -\frac{1}{6} \tilde{x}^3 +\frac{1}{2} \tilde{x}^2 -\frac{1}{3} \tilde{x} \\
	\ell_0(\tilde{x}) &=  \frac{1}{2} (\tilde{x}+1)(\tilde{x}-1)(\tilde{x}-2) \\ \nonumber
        &=  \frac{1}{2} \tilde{x}^3 -            \tilde{x}^2 -\frac{1}{2} \tilde{x} + 1 \\
	\ell_1(\tilde{x}) &= -\frac{1}{2} (\tilde{x}+1)\tilde{x}(\tilde{x}-2) \\ \nonumber
        &= -\frac{1}{2} \tilde{x}^3 +\frac{1}{2} \tilde{x}^2 +            \tilde{x}     \\
	\ell_2(\tilde{x}) &=  \frac{1}{6} (\tilde{x}+1)\tilde{x}(\tilde{x}-1) \\ \nonumber
        &=  \frac{1}{6} \tilde{x}^3                          -\frac{1}{6} \tilde{x}.
\end{align}
\end{subequations}
 The basis may also be viewed as piecewise-defined functions centered at and symmetric about the grid points.  The polynomials then use grid-centered coordinates $\xi = \tilde{x} - x_i$, where $x_i$ is the coordinate of the grid point associated with the polynomial:
\begin{equation}
\label{eq:lagrangegridcenteredbasis}
\ell(\xi) =
\left\{ \begin{array}{r l}
    \frac{1}{2}(|\xi|+1)(|\xi|-1)(|\xi|-2) = \\
    \frac{1}{2} {|\xi|}^3 - {\xi}^2 - \frac{1}{2} {|\xi|} + 1 ,
    & 0 \leq |\xi| < 1, \\
    -\frac{1}{6}(|\xi|-1)(|\xi|-2)(|\xi|-3) = \\
    -\frac{1}{6} {|\xi|}^3 + {\xi}^2 - \frac{11}{6} {|\xi|} + 1, & 1 \leq |\xi| < 2, \\
    0, & |\xi| \geq 2.
       \end{array} \right.
\end{equation}
The first of these equations is obtained by substituting $\tilde{x} = \xi$ in Eq.~(\ref{eq:lagrangebasis}b) or $\tilde{x} = \xi + 1$ in Eq.~(\ref{eq:lagrangebasis}c), and, the second by substituting $\tilde{x} = \xi-1$ in Eq.~(\ref{eq:lagrangebasis}a) or $\tilde{x} = \xi + 2$ in Eq.~(\ref{eq:lagrangebasis}d).  This basis does not have any continuous derivatives across grid points.

\subsection{Piecewise-polynomial-form splines}
Interpolating splines of degree $n$ are piecewise $n^{th}$-order polynomials that reproduce the function values at the grid points. The derivatives of the interpolating splines are continuous up to order $n-1$ across the grid points but do not precisely match those of the function being approximated.  In $d$ dimensions, the implementation of pp-splines that we employ\cite{PrincetonSpline} stores $2^d$ coefficients at each grid point.  These coefficients are the function values and the $2^d-1$ second derivatives.

In one dimension, the cubic pp-spline-represented single-particle orbital is:
\begin{equation}
\label{eq:pp-splineinterpolation1}
\phi_{\rm pp-spl}(\xt) = \sum_{\kappa=1}^{2} \sum_{i=0}^{1} s_i^{\kappa}(\tilde{x}) \; \sigma^{\kappa}_{i}.
\end{equation}
The cubic pp-spline basis polynomials $s_i^{\kappa}$ for uniform grid spacing are\cite{Nord67}:
\begin{subequations}
\label{eq:pp-splinebasis}
\begin{align}
s_0^1(\tilde{x}) &= -\tilde{x} + 1\\
s_0^2(\tilde{x}) &= -\frac{1}{6} \tilde{x}^3 + \frac{1}{2} \tilde{x}^2 - \frac{1}{3} \tilde{x} \\
s_1^1(\tilde{x}) &= \tilde{x} \\
s_1^2(\tilde{x}) &= \frac{1}{6} \tilde{x}^3 - \frac{1}{6} \tilde{x}.
\end{align}
\end{subequations}
In the grid-centered picture, the basis functions are:
\begin{subequations}
	\label{eq:pp-splinegridcenteredbasis}
	\begin{equation}
		s^{1}(\xi) =
		\begin{cases}
			-|\xi| + 1 ,
			& 0 \leq |\xi| < 1, \\
			0, & |\xi| \geq 1
		\end{cases}
	\end{equation}
	\begin{equation}
		s^{2}(\xi) =
		\begin{cases}
			-\frac{1}{6}|\xi|^3 +\frac{1}{2}\xi^2 -\frac{1}{3}|\xi| ,
			& 0 \leq |\xi| < 1, \\
			0, & |\xi| \geq 1.
		\end{cases}
	\end{equation}
\end{subequations}
The cubic pp-spline coefficients $\sigma^{\kappa}$ include the planewave
values $\phi_{\rm PW}$ at the grid points and the constructed
second-derivatives\cite{PrincetonSpline}:
\begin{subequations}
\label{eq:pp-splinecoefficients1}
\begin{align}
	\sigma^{1}_{i} &=       \phi_{\rm PW}(\xt_i) \\
	\sigma^{2}_{i} &= \left.\frac{\partial^2 \phi_{\rm pp-spl}(\xt)}{\partial \xt^2}\right\vert_{\xt_i}
\end{align}
\end{subequations}

The second derivatives are obtained by imposing the condition that the first and second derivatives be continuous at the grid points and two additional boundary conditions, which results in a matrix equation with a diagonally dominant matrix\cite{Nord67}. For periodic boundary conditions, the matrix equation is tridiagonal with corner elements
\begin{eqnarray}
\lefteqn{
\left( \begin{array}{lllcccccc}
        4      & 1      & 0      & 0      & \ldots & 0      & 0      & 0      & 1     \\
        1      & 4      & 1      & 0      & \ldots & 0      & 0      & 0      & 0     \\
        \vdots & \vdots & \vdots & \vdots & \ddots & \vdots & \vdots & \vdots & \vdots\\
        0      & 0      & 0      & 0      & \ldots & 1      & 4      &  1     & 0     \\
        0      & 0      & 0      & 0      & \ldots & 0      & 1      &  4     & 1     \\
        1      & 0      & 0      & 0      & \ldots & 0      & 0      &  1     & 4     \\
\end{array}
\right)
\times
\left(
\begin{array}{l}
        \sigma^{2}(\xt_1)  \\
        \vdots \\
        \sigma^{2}(\xt_i)  \\
        \vdots \\
        \sigma^{2}(\xt_{N_{\rm grid},x})  \\
\end{array}
\right)} \\
&=&
\frac{6}{h^2}
\left(
\begin{array}{l}
        \phi_{\rm PW}(\xt_{N_{{\rm grid},x}}) - 2 \phi_{\rm PW}(\xt_1) + \phi_{\rm PW}(\xt_2)\\
        \phi_{\rm PW}(\xt_1) - 2 \phi_{\rm PW}(\xt_2) + \phi_{\rm PW}(\xt_3)\\
        \vdots\\
        \phi_{\rm PW}(\xt_{i-1}) - 2 \phi_{\rm PW}(\xt_i) + \phi_{\rm PW}(\xt_{i+1})\\
        \vdots\\
        \phi_{\rm PW}(\xt_{N_{{\rm grid},x}-1}) - 2 \phi_{\rm PW}(\xt_{N_{{\rm grid},x}}) + \phi_{\rm PW}(\xt_1)\\
\end{array}
\right)_. \nonumber
\end{eqnarray}

In three dimensions, the cubic pp-spline-represented single-particle orbital is:
\begin{eqnarray}
\label{eq:pp-splineinterpolation3}
\lefteqn{\phi_{\rm pp-spl}(\xt,\yt,\zt) =} \\ \nonumber
   && \sum_{\kappa,\mu,\nu=1}^{2} \sum_{i=0}^{1} s_i^{\kappa}(\tilde{x}) \sum_{j=0}^{1} s_j^{\mu}(\tilde{y}) \sum_{k=0}^{1} s_k^{\nu}(\tilde{z}) \; \sigma^{\kappa,\mu,\nu}_{i,j,k}.
\end{eqnarray}
The trivariate cubic pp-spline coefficients $\sigma^{\kappa,\mu,\nu}$ include the planewave values $\phi_{\rm PW}$ at the knots and the constructed second-derivatives and cross-derivatives\cite{PrincetonSpline}:
\begin{subequations}
\label{eq:pp-splinecoefficients3}
\begin{align}
	\sigma^{1,1,1}_{i,j,k} &=       \phi_{\rm PW}(\xt_i,\yt_i,\zt_i) \\
	\sigma^{2,1,1}_{i,j,k} &= \left.\frac{\partial^2 \phi_{\rm pp-spl}(\xt,\yt,\zt)}{\partial \xt^2}\right\vert_{\xt_i,\yt_i,\zt_i} \\
	\sigma^{1,2,1}_{i,j,k} &= \left.\frac{\partial^2 \phi_{\rm pp-spl}(\xt,\yt,\zt)}{\partial \yt^2} \right\vert_{\xt_i,\yt_i,\zt_i} \\
	\sigma^{1,1,2}_{i,j,k} &= \left.\frac{\partial^2 \phi_{\rm pp-spl}(\xt,\yt,\zt)}{\partial \zt^2} \right\vert_{\xt_i,\yt_i,\zt_i} \\
	\sigma^{2,2,1}_{i,j,k} &= \left.\frac{\partial^4 \phi_{\rm pp-spl}(\xt,\yt,\zt)}{\partial \xt^2 \partial \yt^2} \right\vert_{\xt_i,\yt_i,\zt_i} \\
	\sigma^{1,2,2}_{i,j,k} &= \left.\frac{\partial^4 \phi_{\rm pp-spl}(\xt,\yt,\zt)}{\partial \yt^2 \partial \zt^2} \right\vert_{\xt_i,\yt_i,\zt_i} \\
	\sigma^{2,1,2}_{i,j,k} &= \left.\frac{\partial^4 \phi_{\rm pp-spl}(\xt,\yt,\zt)}{\partial \zt^2 \partial \xt^2} \right\vert_{\xt_i,\yt_i,\zt_i} \\
	\sigma^{2,2,2}_{i,j,k} &= \left.\frac{\partial^6 \phi_{\rm pp-spl}(\xt,\yt,\zt)}{\partial \xt^2 \partial \yt^2 \partial \zt^2}\right\vert_{\xt_i,\yt_i,\zt_i}.
\end{align}
\end{subequations}

\subsection{B-splines}
B-splines are a local basis for splines with one basis function centered at each grid point (or between grid points for even-order functions), such that each basis function is localized and has continuous value and derivative up to some order.  Then, the resulting spline automatically has the same continuity.  In 1-D, an odd-degree B-spline function, $b_i(\xt)$, of degree $n$, is a piece-wise $n^{th}$-order polynomial that is nonzero only in an interval of length $n+1$ and has continuous value and derivatives up to order $n-1$.

The general formula for a B-spline basis polynomial of degree $k$ arises from a recurrence relation\cite{deBoor01}
\begin{equation}
	b_j^k(\xt) = \omega_j^k(\xt)b_j^{k-1} + (1-\omega_{j+1}^k)b_{j+1}^{k-1}(\xt)
\end{equation}
where
\begin{equation}
  \omega_j^k(\xt) = \frac{\xt - \xt_j}{\xt_{j+k}-\xt_j}
\end{equation}
and the B-spline basis polynomial of degree zero associated with grid point $\xt_j$ is a constant $C$ between $\xt_j$ and
$\xt_{j+1}$
\begin{equation}
	b_j^0 =
\begin{cases}
	C, & \xt_j \leq \xt < \xt_{j+1},\\
	0, & \xt < \xt_j \text{ or } \xt \geq \xt_{j+1}.
\end{cases}
\end{equation}
With this definition, the basis function $b_j^k(\xt)$ is nonzero in the interval $(\xt_j,\xt_{j+k+1})$.
Instead, defining $b_j^k(\xt)$ so that $b_j^k(\xt)$ is nonzero in the interval
$(\xt_{j-\lfloor \frac{k+1}{2} \rfloor},\xt_{j+\lfloor \frac{k+2}{2}
\rfloor})$ centers basis $b_j^k(\xt)$ at $\xt_j$ for odd $k$.  Choosing
$C=\frac{3}{2}$ so that the maximum of $b_j^3$ equals one, the cubic ($k=3$)
B-spline basis polynomials for uniform grid spacing are
\begin{subequations}
\label{eq:b-splinebasis}
\begin{align}
	b_{-1}(\tilde{x}) &= -\frac{1}{4} \tilde{x}^3 +\frac{3}{4} \tilde{x}^2 - \frac{3}{4} \tilde{x} + \frac{1}{4} \\
	b_0(\tilde{x})    &=  \frac{3}{4} \tilde{x}^3 -\frac{3}{2} \tilde{x}^2              + 1 \\
	b_1(\tilde{x})    &= -\frac{3}{4} \tilde{x}^3 +\frac{3}{4} \tilde{x}^2 +\frac{3}{4}\tilde{x} + \frac{1}{4} \\
	b_2(\tilde{x})    &=  \frac{1}{4} \tilde{x}^3 .
\end{align}
\end{subequations}
In the grid-centered picture $(b_i(\xt)=b(\xt-\xt_i))$, the basis is\cite{deBoor01}
\begin{equation}
\label{eq:bsplinegridcenteredbasis}
  b(\xi) =
  \begin{cases}
    \frac{3}{4}|\xi|^3 - \frac{3}{2}\xi^2 + 1, &0 \leq |\xi| < 1, \\
    \frac{1}{4}(2 - |\xi|)^3,                  &1 \leq |\xi| < 2, \\
    0,                                         &|\xi| \geq 2.
  \end{cases}
\end{equation}
This basis has continuous first and second derivatives across grid points.

The cubic B-spline-represented single-particle orbital in reduced coordinates is
\begin{equation}
\label{eq:b-splineapproximation}
\phi_{\rm B-spl}(\xt,\yt,\zt) = \sum_{l=-1}^{2} b_l(\tilde{x}) \sum_{m=-1}^{2} b_m(\tilde{y}) \sum_{n=-1}^{2} b_n(\tilde{z}) \; \beta_{lmn}
\end{equation}

\subsubsection{Interpolating B-spline coefficients}
To determine the interpolating B-spline coefficients, we expand the function $\phi_{\rm PW}(x,y,z)$ at each of the grid points in the B-spline basis
\begin{eqnarray}
\lefteqn{\sum_{l=1}^{M_x} \sum_{m=1}^{M_y} \sum_{n=1}^{M_z} B_{il}^x B_{jm}^y B_{kn}^z \; \beta_{lmn} =
\phi_{\rm PW}(\xt_i,\yt_j,\zt_k),} \nonumber \\
&& 1 \le i \le M_x, \;\; 1 \le j \le M_y, \;\; 1 \le n \le M_z
\end{eqnarray}
$B_{il}^x B_{jm}^y B_{kn}^z = b_l(\xt_i) b_m(\yt_j) b_n(\zt_k)$ is the value at $(\xt_i,\yt_j,\zt_k)$
of the basis function centered at grid point at $(\xt_i,\yt_j,\zt_k)$,
and $M_i$ is the number of grid points in the $i^{th}$ direction.
Since a cubic B-spline basis function is nonzero at only 3 grid points along the direction of each lattice vector the matrix $\bf B$ is triadic.
For periodic or antiperiodic boundary conditions, $\bf B$ is tridiagonal with corner elements.
Solving these equations in three stages\cite{Esler_unpub} produces the B-spline coefficients,
$\beta_{lmn}$, at a computational cost of ${\it O}(M_x M_y M_z)$,  which is negligible compared to the cost of the QMC calculation.
An advantage of interpolating B-splines\cite{Einspline} over smoothing B-splines
is that interpolations do not require an evenly-spaced grid.

\subsubsection{Smoothing B-spline coefficients}
\label{smoothing_spline}

Instead of choosing the B-spline coefficients to construct an interpolating approximation, an alternative is to choose them such that the
Fourier components of the approximation exactly match the nonzero components of the planewave expansion.
This gives
\begin{equation}
\label{eq:b-splinecoefficients}
        \beta_{lmn} = \sum_{\vector{G}} \frac{c_{\vector{G}}}{\gamma_{\vector{G}}} \exp(\imath \left[\vector{G} \cdot (x_l,y_m,z_n)\right])
\end{equation}
where $\gamma_{\vector{G}}$ is the 3D Fourier transform of an individual basis spline\cite{Hernandez97,Alfe04}:
\begin{eqnarray}
  \gamma_{\vector{G}} &=& \prod_{m = \{x,y,z\}} \frac{3 \left[3 - 4 \cos(G_{m}) + \cos(2 G_{m})\right]}{{G_{m}}^4} \nonumber \\
  &=& 24 \prod_{m = \{x,y,z\}} \left(\frac{\sin(G_{m}/2))}{G_{m}} \right)^4 \nonumber \\
  &=& \frac{3}{2} \prod_{m = \{x,y,z\}} {\rm sinc}^4\left(\frac{G_{m}}{2}\right).
\end{eqnarray}
Fourier expanding the B-spline representation of Eq.~(\ref{eq:b-splineapproximation}) with the choice of $\beta_{lmn}$ given
in Eq.~(\ref{eq:b-splinecoefficients}) yields Fourier components that exactly match the nonzero components of the planewave expansion.
However, the B-spline has additional higher frequency components that are very small in magnitude.

\end{document}